\documentclass[twocolumn,amsmath,amssymb,prb,superscriptaddress,longbibliography,nobalancelastpage]{revtex4-2}
\UseRawInputEncoding
\usepackage{titlesec}
\usepackage[colorlinks,bookmarks=false,citecolor=blue,linkcolor=blue,urlcolor=blue]{hyperref}
\usepackage{epsfig}
\usepackage{color}
\usepackage{mathtools}
\usepackage{subfigure}
\usepackage{multirow}
\usepackage{rotating}
\usepackage{lipsum}
\usepackage{graphicx} 
\usepackage{dcolumn}  
\usepackage{mwe}
\input{insbox}
\usepackage[titletoc,title]{appendix}
\usepackage[graphicx]{realboxes}
\usepackage{tabularx}
\usepackage{bm}
        {\pagebreak[4]\global\pdfpageattr\expandafter{\the\pdfpageattr/Rotate 0}}%

\newcommand{\be}{\begin{equation}}
\newcommand{\ee}{\end{equation}}

\begin{document}

\title{A majority gate for two-dimensional ferromagnets lacking inversion symmetry}

\author{Nikolaos Ntallis}
\email{nikos.ntallis@physics.uu.se}
\affiliation{Department of Physics and Astronomy, Uppsala University, Box 516, SE-75120 Uppsala, Sweden}
\author{Anders Bergman}
\affiliation{Department of Physics and Astronomy, Uppsala University, Box 516, SE-75120 Uppsala, Sweden}
\author{Anna Delin}
\affiliation{Department of Applied Physics, School of Engineering Sciences, KTH Royal Institute of Technology, AlbaNova University Center, SE-10691 Stockholm, Sweden}
\affiliation{Swedish e-Science Research Center (SeRC), KTH Royal Institute of Technology, SE-10044 Stockholm, Sweden}
\author{Olle Eriksson}
\affiliation{Department of Physics and Astronomy, Uppsala University, Box 516, SE-75120 Uppsala, Sweden}
\affiliation{School of Science and Technology, Örebro University, Fakultetsgatan 1, SE-701 82 Örebro, Sweden}
\author{Erik Sj\"oqvist}
\affiliation{Department of Physics and Astronomy, Uppsala University, Box 516, SE-75120 Uppsala, Sweden}
\author{Danny Thonig}
\affiliation{School of Science and Technology, Örebro University, Fakultetsgatan 1, SE-701 82 Örebro, Sweden}
\affiliation{Department of Physics and Astronomy, Uppsala University, Box 516, SE-75120 Uppsala, Sweden}
\author{Manuel Pereiro}
\email{manuel.pereiro@physics.uu.se}
\affiliation{Department of Physics and Astronomy, Uppsala University, Box 516, SE-75120 Uppsala, Sweden}

\begin{abstract}
The manipulation of topologically protected field configurations, already predicted and experimentally observed in non-centrosymmetric magnets, as skyrmions, merons and antimerons could definitely have potential applications in logic gate operations as carriers of information. Here, we present and elaborate a proof of concept on how to construct a three-input non-canonical majority gate on a kagome ferromagnet lacking inversion symmetry. By taking advantage of the existence of edge modes in a Kagome magnet, it is possible to create topological excitations as merons and antimerons at the edge of the material. Using atomistic spin dynamics simulations, we determine the precise physical conditions for the creation and annihilation of merons and antimerons and, in a second stage, we describe  the majority gate functionality.

\end{abstract}

\date\today
\maketitle

\section{INTRODUCTION}
Conventional microelectronic integrated circuits (ICs) work by controlling the flow of electrons through transistor switches.  In addition to electrical charge, electrons have another degree of freedom given by the spin, thus permitting alternate representations of binary digits. This fact is key for a new technology named as spintronics, in which the spin and the charge of the electron are used  to represent bits and carry out data processing. In general, this technique promises lower-powered, higher-speed and nonvolatile devices \cite{spintronics1}. A basic logic gate, such as AND or NOT, is an idealized model of computation or physical electronic device implementing a Boolean function, that is, a logical operation performed on one or more binary inputs that produces a single binary output. Thus, a logical gate allows the digital IC to combine numbers from memory in arithmetic calculations. The aforementioned gates are based on the concept of comparing two input channels (with values 0 or 1) and produce an output (with values 0 or 1) depending on the operation. 

More complex operations involve the so-called majority gate. In contrast to basic gates, the majority gate has a number of inputs bigger than 2 and the output is governed by the majority of  its inputs. Thus, if  more than 50\% of the input is 1 then it  will produce 1, otherwise zero \cite{majority}. Recently, the research in spintronics has accelerated in an effort to find an alternative to the existing complementary metal–oxide–semiconductor-based electronics. Several physical phenomena have been exploited to develop novel spin logic devices \cite{spinlogic1,spinlogic2}. Their main concepts are based on the manipulation of domain-walls \cite{walls1,walls2,walls3}, but it is has been addressed that these magnetic nanostructures are affected  by pinning due to material defects \cite{defects}. More recently, other spin structures, known as skyrmions, have been actively and intensively studied \cite {skya1}. Magnetic skyrmions \cite{skya2,skyb1,skyb2,skyb3,skyb4,skyb5,skyb6} are chiral spin configurations that can be found either in magnetic materials with a non-centrosymmetric crystal structure or in  ultrathin films in which the inversion symmetry is broken by the presence of non-equivalent interfaces. Skyrmion states are generally present in systems lacking inversion symmetry and with non-zero Dzyaloshinskii–Moriya interactions~\cite{skyb6,skyc1,skyc2,skyc3} (DMIs). Because of their topological nature, the magnetic skyrmions are argued to be protected, and consequently  stable, similarly to the other skyrmionic configurations first predicted in elementary particle physics~\cite{c1}. These configurations were later found in other fields of physics such as magnetism, liquid crystals~\cite{c2} or Bose–Einstein condensates~\cite{c3}. 

A magnetic skyrmion is argued to be topologically stable because it carries a topological number~\cite{topo1,topo2}. 
The skyrmion can be destroyed or created at the edges of structures and can be manipulated by application of external fields or currents~\cite{skyedge1,skyedge2}. Skyrmions have proven to be more robust to pinning compared to domain walls \cite{skya2} and based on these properties, several skyrmion-based logic devices~\cite{skydev1,skydev2,skydev3} have been proposed. The skyrmion velocity in these devices is limited by its transverse displacement due to the Magnus force~\cite{magnus1}. In order to avoid the latter effect, antiferromagnetically coupled magnetic bilayer systems have been proposed, where the skyrmions are nucleated in each of these layers~\cite{magnus2}, such that, the Magnus force cancels out. However, in such cases, engineering of the materials is of great importance in order to obtain  perfect coupling of the skyrmions. On top on that, inside one material/structure, one can have only one type of skyrmion. Thus, a majority gate can be constructed  only in a implicit sense, i.e., two channels must first interact with each other independently, and then, the output result must interact with a third channel in order to produce the final output. 

In addition to the discovery of skyrmions, there has also been a great effort in searching for new forms of topological structures.  One promising structure is the so-called meron, which was originally described in the context of quark confinement ~\cite{meron,meron1} and identified in condensed matter physics, as a magnetic vortex and topologically equivalent to one-half of a skyrmion. Merons have been mainly studied in magnetized film systems, as they are more accessible experimentaly   due to the intrinsic demagnetization field ~\cite{meron2}.
Individual merons have been reported to exist  only  in confined geometries ~\cite{meron2,meron3,meron4} . In a system film, merons must exist in pairs (meron and it’s counterpart antimeron) or groups~\cite{meron5,meron6}. Experimentally, multiple vortices were observed  as transient states \cite{meron7,meron8} or in aggregated groups ~\cite{meron9}.
Recently, the  topologically non-trivial meron lattice was observed in the chiral magnet Co$_8$Zn$_9$Mn$_3$ ~\cite{meron10} in the form of square lattices which transformation into a hexagonal lattice of skyrmions in the presence of a magnetic field at room temperature. Moreover,  observations with decreasing temperature reveal that the square lattice of merons and antimerons relaxes to non-topological in-plane spin helices, highlighting the different topological stability of merons, antimerons and skyrmions.
For the aforementioned reasons, we turn our attention to the kagome ferromagnet. For a complete definition of a kagome ferromagnet see Ref.~\cite{pereiro}. As a crystal structure, the kagome lattice is composed of two triangular lattices, one with lattice constant $a$ and the other with lattice constant $2a$, so that, there exist large empty hexagonal spaces in the crystal geometry. 
Skyrmions and merons are appealing for applications in information storage or logic devices. However, they need to meet several requirements in order to be useful for magnetic applications, namely, they need to have high mobility, small size and should allow for full control of the direction of movement. In this article, we report on the conditions under which controlled topological structures (with special attention to merons and antimerons) are created in a kagome lattice, and we show that this system has very interesting technologically relevant properties.

The rest of the article is organized as follows: In Sec.~\ref{theoretical_model}, we introduce the theoretical model together with the coupling parameters used in the method. After establishing the physical conditions for controlling the creation and annihilation of merons and antimerons in Sec.~\ref{meron-antimeron}, we describe in detail the majority gate functionality of a 2D kagome magnet in Sec.~\ref{device}. The article concludes with a brief summary and discussion of the reported results in Sec.~\ref{conclusions}.


\section{THEORETICAL MODEL}
\label{theoretical_model}

Our theoretical modelling is based on the following hamiltonian:
\begin{equation}
	\mathcal{H}=-\sum_{i,j}\left[J_{ij}\bm{s}_i\bm{s}_j+\bm{D}_{ij}\bm{s}_i \times \bm{s}_j\right]-g \mu_B  \sum_i \bm{s}_i \bm{B}_i
\end{equation}

\noindent where $i, j$ denote first-neighbour indices,  $\bm{s}_i$ is the atomic magnetic moment with $|\bm{s}_i|=2$ $\mu_{B}$, $J_{ij}$ is the exchange interaction and $\bm{D}_{ij}$ is the vector form of DM interaction. The last term represents the Zeeman term under the influence of a local external magnetic field $\bm{B}_i$. Here, we assume that the Heisenberg exchange and Dzyaloshinskii-Moriya interaction to be the same for every interacting pair of atoms, so that, $J_{ij}=J$ and $\lvert \bm{D}_{ij}\rvert=D$. 

The relevant factor in our analysis is represented by the ratio between the antisymmetric ($D$) and isotropic ($J$) exchange interactions, namely $D/J$. This ratio captures the response of our simulated system. In experiments \cite{ratio1}, for the Lu$_2$V$_2$O$_7$ vanadate with kagome planes along the [111] direction, the ratio  $D/J$ has been found to be as high as 0.32. In the calculations shown on this work, and in order to mimic realistic materials, $J$ was set constant to be 1 mRy while $D$ was varied from 0.1~mRy to 0.4~mRy in steps of 0.1~mRy.

In order to capture the dynamical properties of spin systems at finite temperatures, we used an atomistic spin dynamics (ASD) approach (as described e.g. in Ref.~\cite{olle}). The equation of motion of the classical atomistic spins at finite temperature is governed by the Langevin dynamics via a stochastic differential equation, normally referred to as the atomistic Landau-Lifshitz-Gilbert equations, which can be written in the form:

\begin{eqnarray}
	\frac{\partial \bm{s}_i}{\partial t}=&-&\frac{\gamma}{1+\alpha^2}\bm{s}_i\times\left[\bm{B}_i+\bm{b}_i\right]\nonumber\\
	&-&\frac{\gamma \alpha}{(1+\alpha^2)|\bm{s}_i|}\bm{s}_i\times\left(\bm{s}_i\times[\bm{B}_i+\bm{b}_i]\right)
\end{eqnarray}

\noindent where $\gamma$ is the gyromagnetic ratio and $\alpha_i$ denotes the site-depended damping. The effective field $ \bm{B}_i=-\partial{\mathcal{H}}/\partial \bm{s}_i $ acts on each site and the stochastic field $\bm{b}_i$ introduces thermal fluctuation effects at a temperature $T$. The latter is assumed to be a Gaussian random process obeying the following equations:
\begin{eqnarray}
	\langle b_{i,m}(t)\rangle&=&0\nonumber\\
	\langle b_{i,m}(t)b_{j,n}(t^\prime)\rangle&=&\frac{2 \alpha k_B T }{|\bm{s}_i|\gamma}\delta_{ij}\delta_{mn}\delta(t-t^\prime)
	\label{stochastic}
\end{eqnarray}

\noindent where $i, j$ represent site indices and $m, n$ represent Cartesian coordinates of the field and $k_B$ is the Boltzmann constant. 

The skyrmion number represents a topological index of the field configurations and is evaluated here as:

\begin{equation}
	\mathfrak{R}=\frac{1}{4\pi}\int_\mathcal{S}\bm{\hat{s}}\cdot\left(\frac{\partial \bm{\hat{s}}}{\partial x}\times \frac{\partial \bm{\hat{s}}}{\partial y}\right)dxdy
	\label{skyrmion_number}
\end{equation}

\noindent where $\hat{\bm{s}}$ is the unit vector of the local magnetization and $\mathcal{S}$ is the surface surrounding the size of the skyrmion. In this work, the integration surface $\mathcal{S}$ is defined as a region with 10x10 unit cells  with the {\it central cell} defined as the one in which we found the lowest negative value of magnetization's z-component, which in general is expected to be located the center of a meron/antimeron particle. The sign of the z-component of the magnetization in the {\it central cell} determines if the particle is a meron (negative sign) or an antimeron (positive sign).

\begin{figure}[t]
 \centering
 \includegraphics[width=0.48\textwidth]{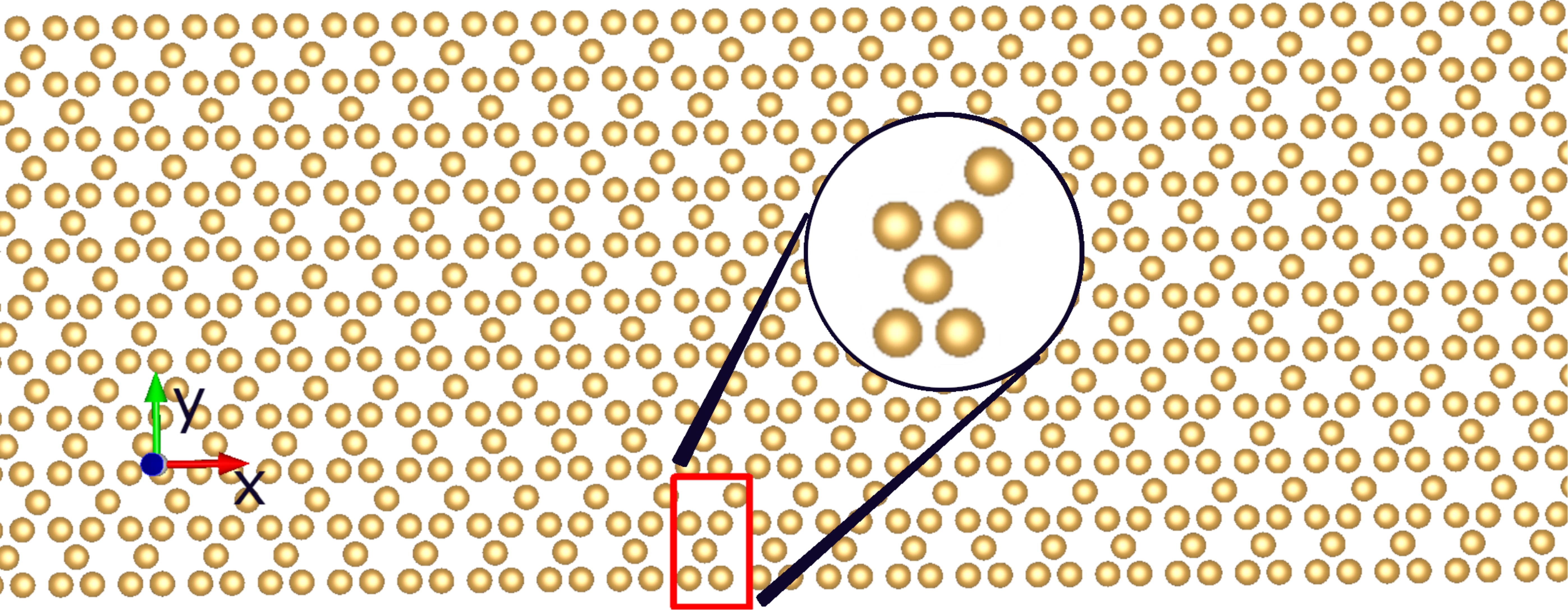}
 \caption{Fragment of the simulation cell in the shape of a kagome lattice. The box highlighted in red represents the unit cell used in this article and it is composed of 6 atoms, as shown in the magnified section.}
 \label{fig1(a)}
\end{figure}

We apply here atomistic spin dynamics calculations on a monolayer of 400x40 cells of the ferromagnetic kagome lattice. In Fig.~\ref{fig1(a)}, a portion of the simulation domain is shown with the unit cell framed by a rectangular red region. We have used 6 atoms per unit cell with the aim to describe the system with orthogonal unit vectors along x- and y-Cartesian axis.
As the Kagome lattice lacks of inversion symmetry a non zero DM can exist. The DM vectors have components normal and perpendicular to the kagome plane, as shown in Fig.~1 in supplementary information, the kagome structure inducing a frustration of the DM interaction \cite{pereiro}.


Before performing ASD simulations, the ground state of the system was found by using an annealing protocol, applied under a magnetic field of 0.1 T  in the whole sample for ruling out any degenerate state at zero external magnetic field. Under these conditions, the ground state of the system is non-collinear but almost a ferromagnetic state. The word "almost" refers here to the fact that a tilting of the moment appears due to the DM interaction, which is enhanced by increasing the DM strength. See supplementary information for further details.

A local external magnetic field is applied parallel to the surface normal of the lattice, in a region with dimension $x_c$x$y_c$ cells, which induces an excitation of the system.  The size of this perturbation is important as for very small excitations, smaller than 11x4 cells, no merons were found to be created. On the other hand, for excitations with large size, larger than 16x7 cells the most frequent solution was the creation of a meron/antimeron pair. The optimal size of the excitation cell for the stable creation of merons or antimerons was found to be 13x6 cells and 12x5 cells for $D<0.25$~mRy and $D>0.25$~mRy, respectively. It has to be noted here that also different sizes of the excitation regions were used, so that the same concept also applies for semi-circular excitation shapes. For the sake of simplicity, we here discuss only results from excitations that had a shape from $x_c$x$y_c$ cells. 

\section{CONTROLLING THE CREATION OF MERONS AND ANTIMERONS}
\label{meron-antimeron}
In this work, we present an idea of creating a majority gate on a single material. For this reason, we focus our study on a kagome ferromagnet. It has been demonstrated that in a kagome ferromagnet edge modes exist leading to the formation of  meron-antimeron pairs by applying external excitations on the edge~\cite{pereiro}. The meron-antimeron particle stabilization depends on the balance between $J$ and $D$. As merons and antimerons are particle and antiparticle solutions, they possess opposite momentum and, therefore they move in opposite directions on the same edge of the kagome lattice. For the same reason, these two structures annihilate each other in a collision event. Thus, in a kagome lattice, we have the opportunity to  create two different bits, the meron and the antimeron, and there exist an internal procedure (annihilation)  that these two objects can be compared, as they move in different directions. These properties, alongside the size and direction of D with respect to J, makes a magnetic material sustaining a kagome lattice a candidate for a majority gate functionality, as it is anticipated in Sec.~\ref{conclusions}. 
For the gate to be functional, a controllable way of creating isolated meron/antimerons is mandatory. We consider here magnetic excitations of a 2D kagome lattice, by applying a local torque. For the sake of simplicity, the local torque is produced by a local magnetic field normal to the kagome strip. We assume a very low temperature (T=1 mK) in the majority of our simulations, unless stated otherwise. The choice of low temperature was made since the simulated results are easier to be observed then. This does not prevent the functionality of the proposed system to be operative also at higher temperatures. By allowing an external field to act locally at different times, $t_h$, we can create isolated merons/antimerons on the edges of the kagome stripe. 

\begin{figure}[t]
 \centering
 \includegraphics[width=0.52\textwidth]{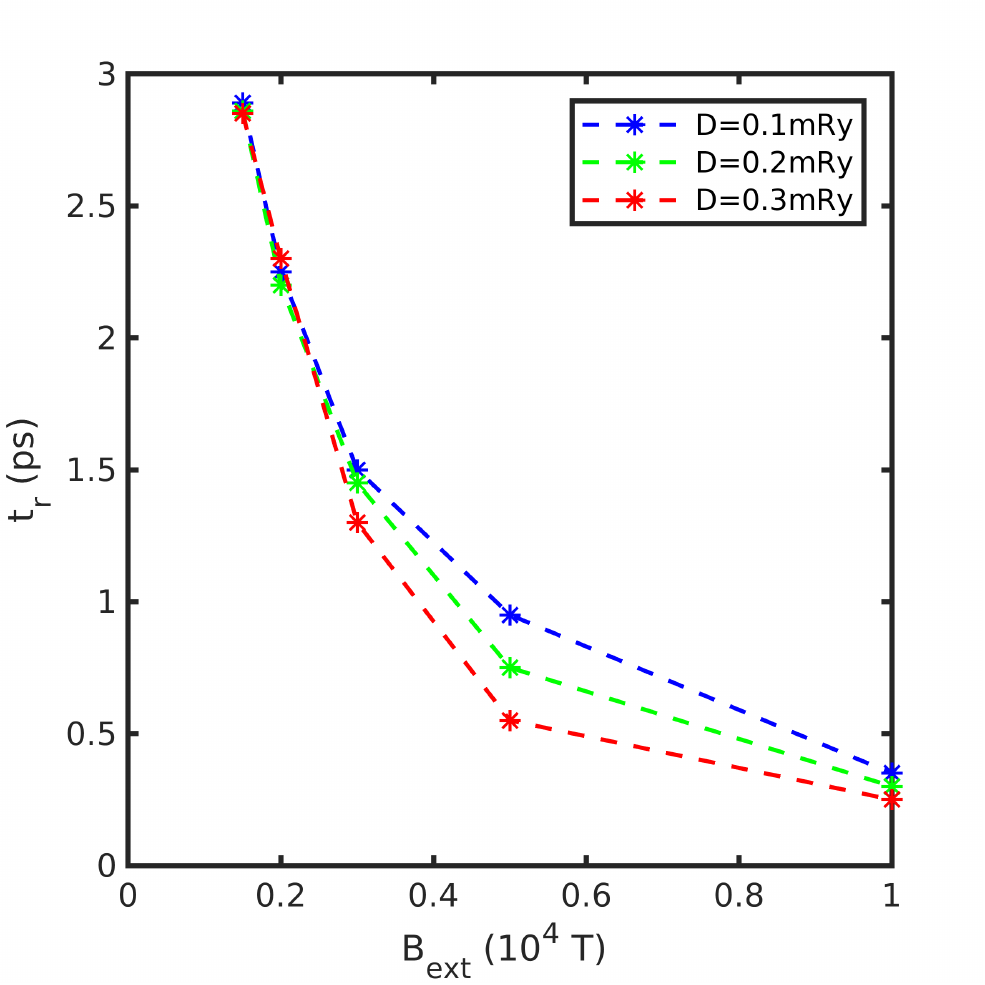}
 \caption{Simulated reversal time, $t_r$, plotted with respect to the localized external field applied on the excited region for different strengths D of the Dzyaloshinskii–Moriya interaction. The other parameters used in this simulation are $J=0.1$~mRy and $\alpha=0.01$.}
 \label{fig2}
\end{figure}

The excitation time, $t_h$, cannot be smaller than the reversal time, $t_r$, the latter being defined as the time needed to reverse the magnetization in the initial region of excitation. For the extraction of $t_r$, we apply the local field and measure the time at which the mean moment value of the excited region reaches the value of $-0.8M_S$, where  $M_S$ is the mean initial moment of the excited region. Since a very tiny and local region needs to be reversed, a high and very localized external field is needed to break the atomic exchange interactions. Thus, we need to identify a proper field/time pair to achieve the desired excitation. Figure~\ref{fig2} shows the  variation of $t_r$  with respect to the external field strength for different values of $D$. In our simulations  $t_h$  ranges in the interval $[1.5,2.5]$ ps.

As shown in Figure~\ref{fig2}, increasing $D$ causes the reversal time to decrease as the non-collinearity of the magnetic structure is enhanced and, consequently, the atomistic isotropic interactions are softened. For the selected values ($J=1$~mRy, $D=0.1-0.4$mRy), a field of 1500 T is needed to achieve local reversal in a time close to 3 ps.  With the aim to shorten the simulation time, we applied magnetic fields of the order of $10^4$ T in the rectangular region for the majority of the simulations presented in this work.

\begin{figure}[t]
\includegraphics[scale=1.2,width=0.48\textwidth]{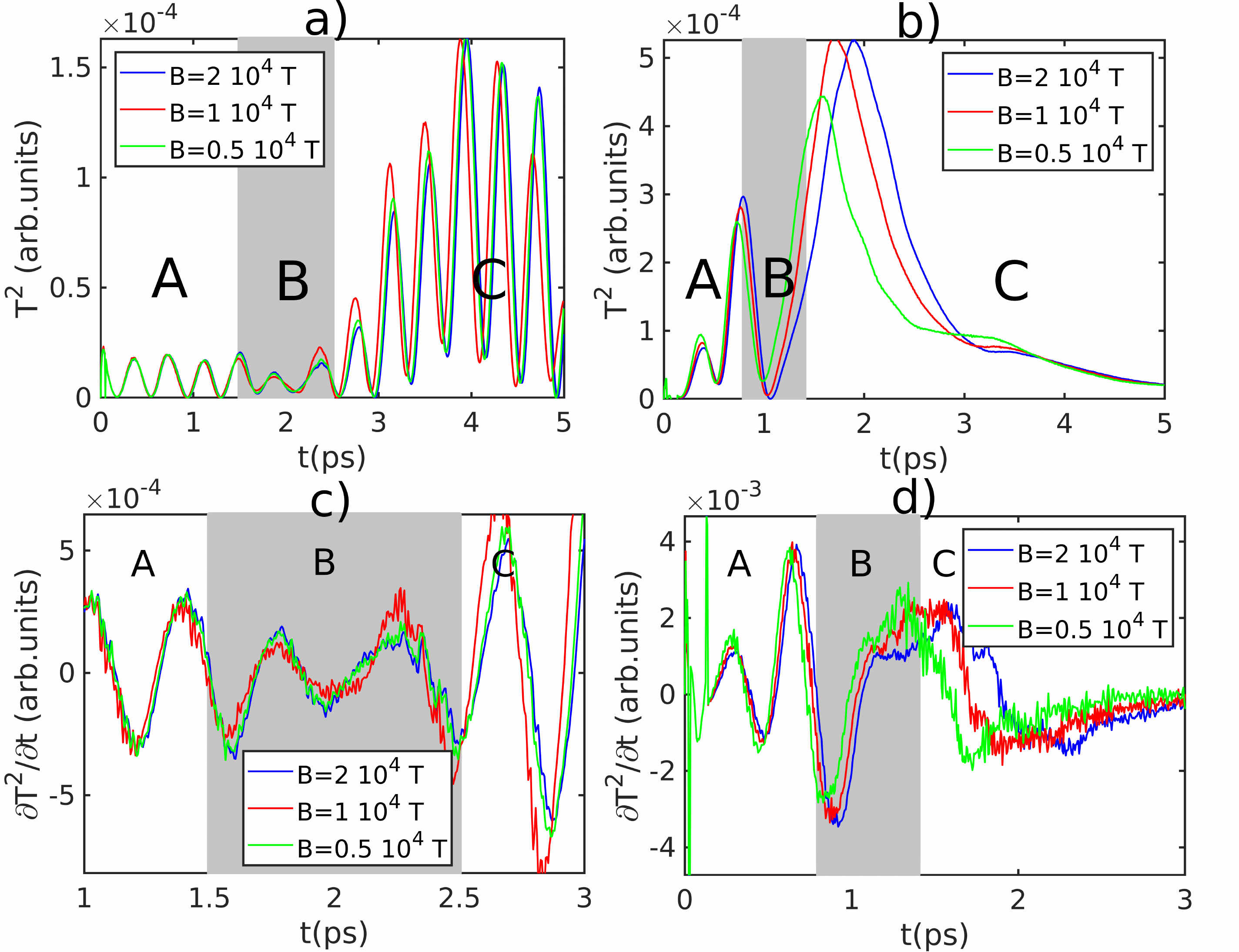}
\caption{Torque (for definition see text) squared as a function of time for different values of the external field. Here $J=1$~mRy, $D=0.1$~mRy and a) $\alpha=0.01$ b) $\alpha=0.06$. $(\partial T^2/\partial t)$  for c) $\alpha=0.01$ d) $\alpha=0.06$. The  region  A  is  a  transition state  involving  dynamics,  the region  B  is the one where meron and antimeron creation is achieved, and region C refers and an almost coherent rotation of the excited region.
}
\label{fig3}
\end{figure}

The highly nonlinear process  of the meron/antimeron creation makes the construction of a solid model extremely difficult for predicting with high accuracy if the outcome of the excitation in the rectangular region will be either a meron or an antimeron. However, by tracing the torque variation when the external field is on, we can identify a time window were always either single merons or antimerons are created.  Figures~\ref{fig3} a) and b) show the variation of the in-plane magnitude of the torque given by:

\begin{equation}
	T^2=\int_\mathcal{S} \left(\frac{\partial s_x}{\partial t}\right)^2+ \left(\frac{\partial s_y}{\partial t}\right)^2
	\label{torque1}
\end{equation}

\noindent with respect to time for different applied field strengths with $\alpha=0.01$ and $\alpha=0.06$, respectively. 
The integration is performed in a rectangular region ($\mathcal{S}$) which embrace the excitation region together with a frame of one unit cell width along each direction of the rectangular region. In this region, we only consider the in-plane components of the magnetic moment ($s_x, s_y$). As commented above, the creation of isolated merons and antimerons is mainly produced in a region of at least 13x6 unit cells or 12x5 cells for $D<0.25$ mRy and $D>0.25$ mRy, respectively. By increasing the excitation size, the meron-antimeron pair production is more pronounced, and consequently, this possibility needs to be ruled out for the sake of having a good majority gate functionality with isolated merons and antimerons.

\begin{figure}[t]
 \centering
 \includegraphics[width=0.52\textwidth]{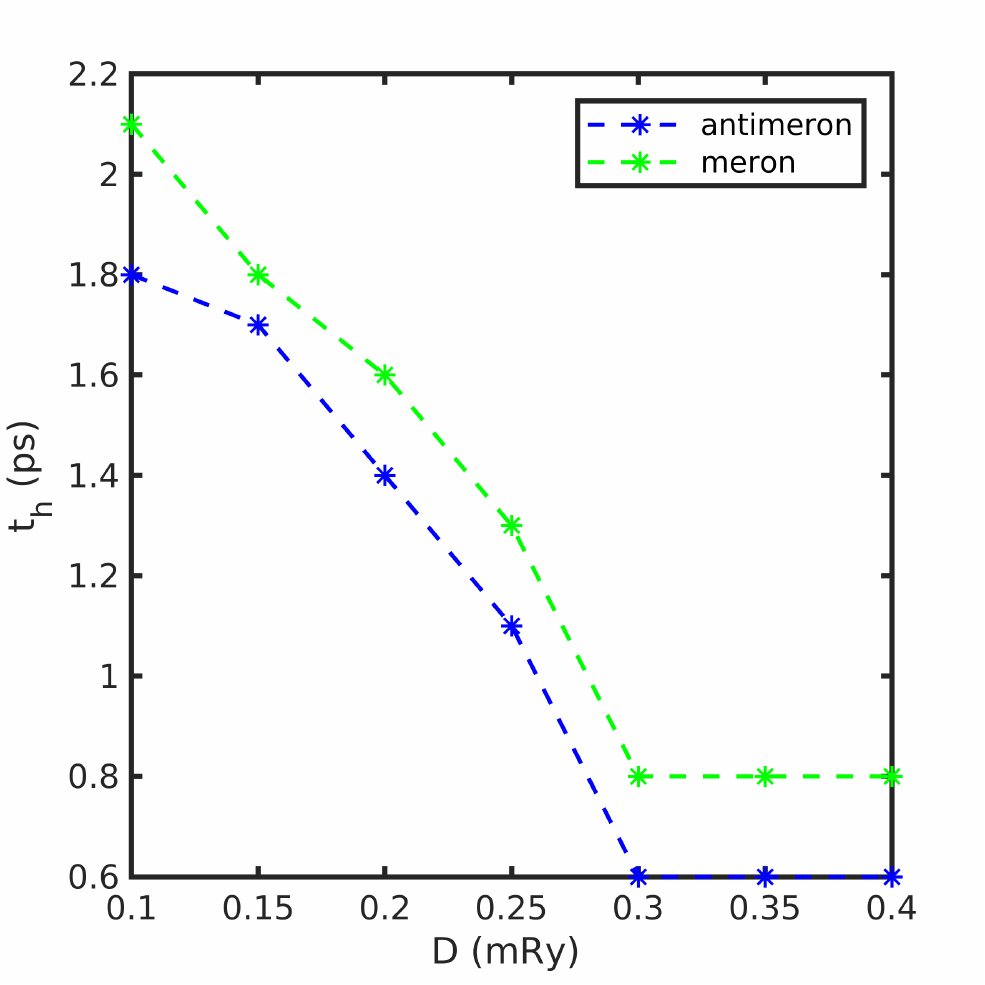}
 \caption{Time $t_h$ needed for the creation of a meron and antimeron as a function of D. Here the external field is $10^4$ T and  $\alpha$=0.01. }
 \label{fig4}
\end{figure}

In Figs.~\ref{fig3}a)-b), we can identify three different regions of interest labelled as regions A, B and C, respectively. The region A (left from shaded region) is a transition state involving dynamics due to the application of the external field. Region C (right from the shaded region) refers to a temporal interval where the external torque has been applied for such time that the first neighborhood around the excitation performs almost coherent rotation with respect to the external field and thus the reversed region.  This region can be clearly identified by the large increment of the integrated torque. Last, the region B (shaded region), i.e. the intermediate region between regions A and C, is a temporal interval where  isolated merons or antimerons can be created. In region B, the creation of merons or antimerons is identified by a change of sign of the slope of $(\partial T^2/\partial t)$ with respect to the excitation time ~\ref{fig3}c)-d). Inside the region B, a positive slope produces an antimeron whereas a negative slope a meron. Transition from positive to negative slope is always done by passing a critical point. Passing through this critical point interchanges the torque components that favor clock-wise (antimeron) or counterclock-wise (meron) chilarity. In the vast majority of our simulations, the critical point between a sign change of the slope, generally provides a coupled meron-antimeron created on the edge and moving along the bulk of the material. The shape of $T^2$ remains the same as the DM interaction increases. It has to be noted here though that the increment of the DM interaction affects the time the external field needs to be switched off. As shown in Fig.~\ref{fig4}, the time $t_h$ required to create a meron or an antimeron reduces  with respect to the DM strength as the non-collinearity is enhanced, and consequently, this fact leads to a faster realization of single merons and antimerons. Interestingly, the $t_h$ becomes almost constant after $D>0.3$~mRy indicating that for large $D/J$ ratios, the time required to generate merons or antimerons tends to be independent of the DM interaction and, instead, depends on the applied magnetic field and the damping.


For the same material parameters and same excitation field, the meron/antimeron window becomes narrower (cf. Fig.~\ref{fig3}) by increasing the damping parameter and also $t_h$ becomes smaller. At a higher damping, the magnetization is synchronized with the external field faster as oscillating modes are damped out. Even though a higher damping seems useful for an easier identification of region B since there are less changes in sign of the torque and the signal of $T^2$ is more pronounced, it has the downside that the increase of the damping reduces the lifetime of a meron. 

\begin{figure}[t]
 \centering
 \includegraphics[width=0.48\textwidth]{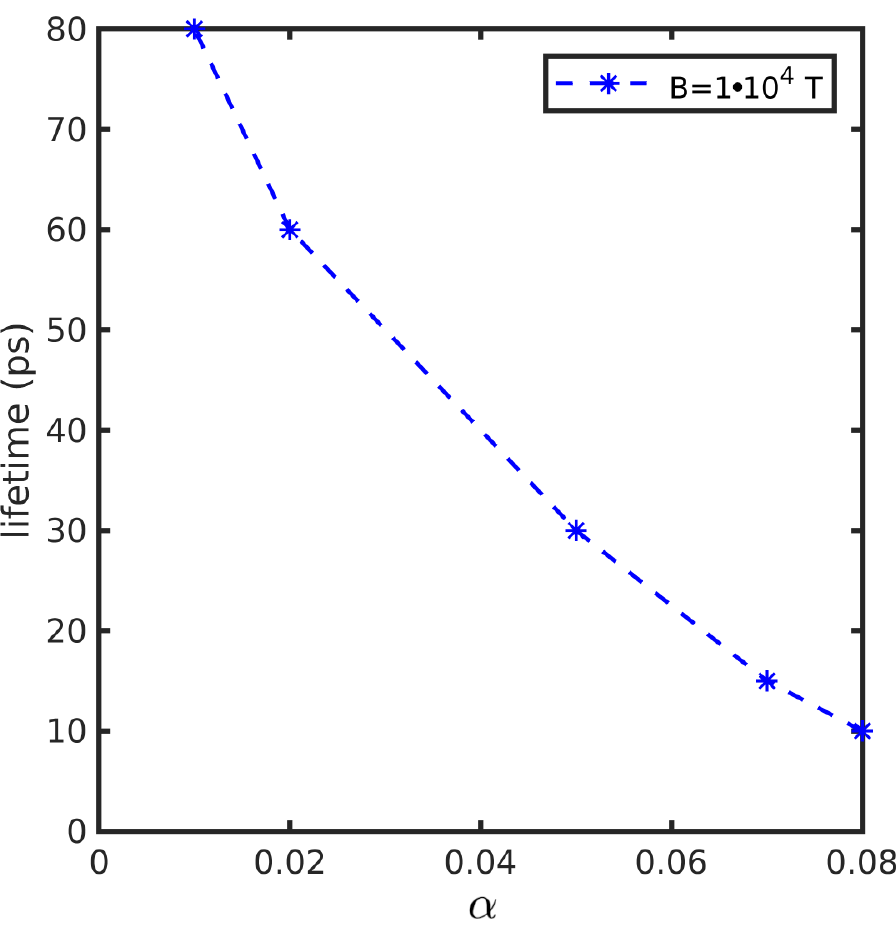}
 \caption{Lifetime of meron with respect to the damping parameter.}
 \label{fig5}
\end{figure}

Figure~\ref{fig5} shows the calculated lifetime of a meron for different values of the damping parameter. Lifetime of merons is important as it is critical for the response time of a device. In general, we would like this parameter to be infinite so that the device response would be unconstrained. Above a value of $\alpha=0.08$~merons/antimerons cannot be created at all, indicating an extreme sensitivity on dissipation as they are not static solutions. Reducing the damping, the lifetime increases and in the asymptotic limit tends to infinity when $\alpha\to 0$. The criterion we consider to accept a solution (meron/antimeron) to produce logical operations is for those particles that have a lifetime longer than 60~ps. In terms of damping, the criterion to accept the particles is for $\alpha \lesssim 0.02$. We note that this is higher than the experimentally observed damping for most ferromagnets, which demonstrates that the damping is not a limiting factor for choosing materials to be used in the skyrmionic majority gate suggested here.


\begin{figure}[t]
 \includegraphics[width=0.48\textwidth]{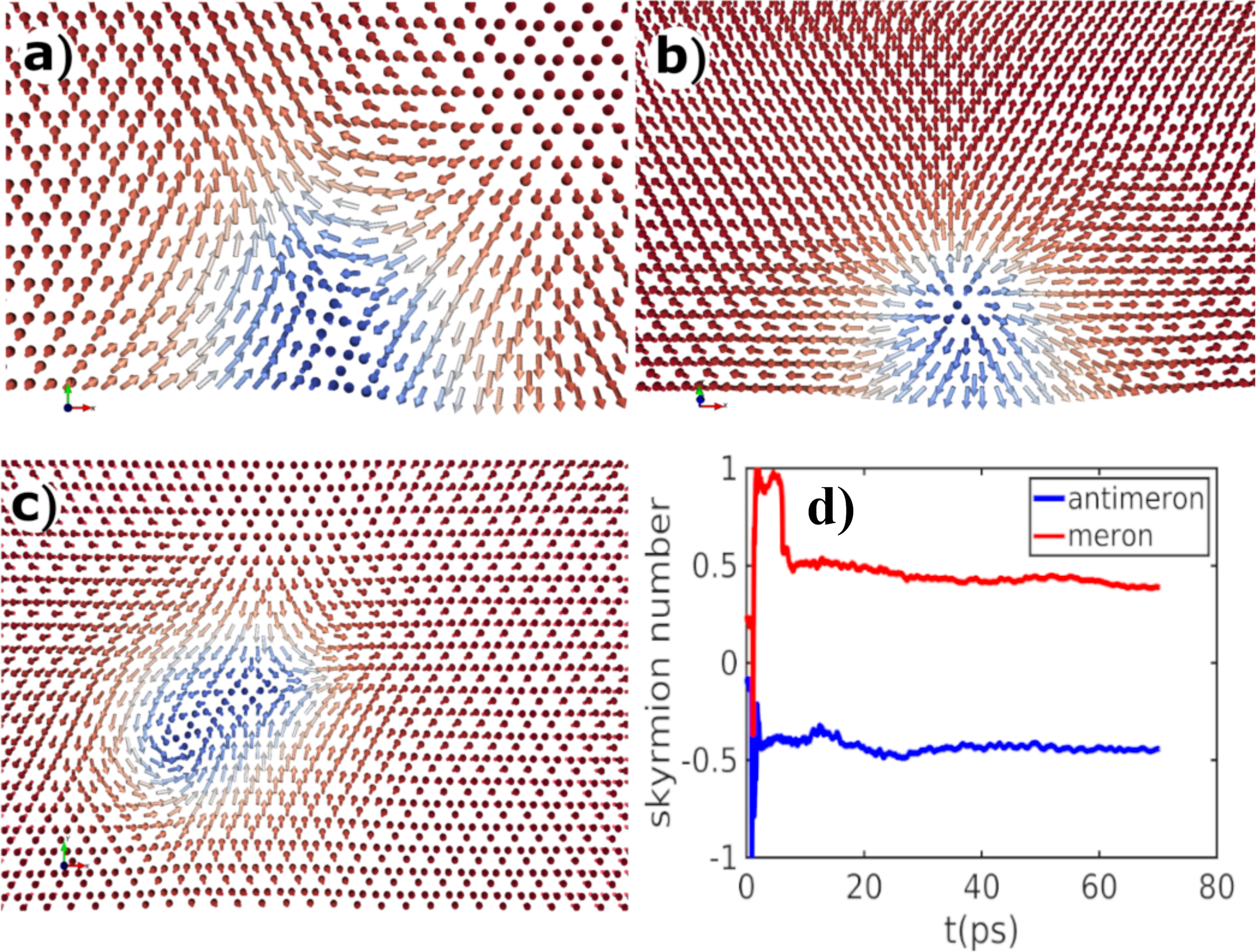}
 \caption{Structure of a) antimeron, b) meron, and c) meron-antimeron pair. d) The variation of the skyrmion number for the first 70 ps after the creation of an antimeron (blue) and a meron (red). The exchange parameters are $J=1$~mRy and $D=0.1$~mRy. 
 }
 \label{fig6}
\end{figure}

Figure~\ref{fig6} shows calculated structures of an antimeron (cf. Fig.~\ref{fig6}(a)), meron (cf. Fig.~\ref{fig6}(b)), and a meron-antimeron pair (cf. Fig.~\ref{fig6}(c)) created with the application of an external torque as described previously. Apart from the visual identification, one can also identify these topological objects, by the calculation of the skyrmion number (see Fig.~\ref{fig6}(d)). Notice that due to the fully atomistic nature of the present model, the skyrmion number is not necessarily an integer number, as would be expected for a skyrmion at the micromagnetic scale where the magnetization is a continuously differentiable function of the position.

\begin{figure}[t]
 \includegraphics[width=0.48\textwidth]{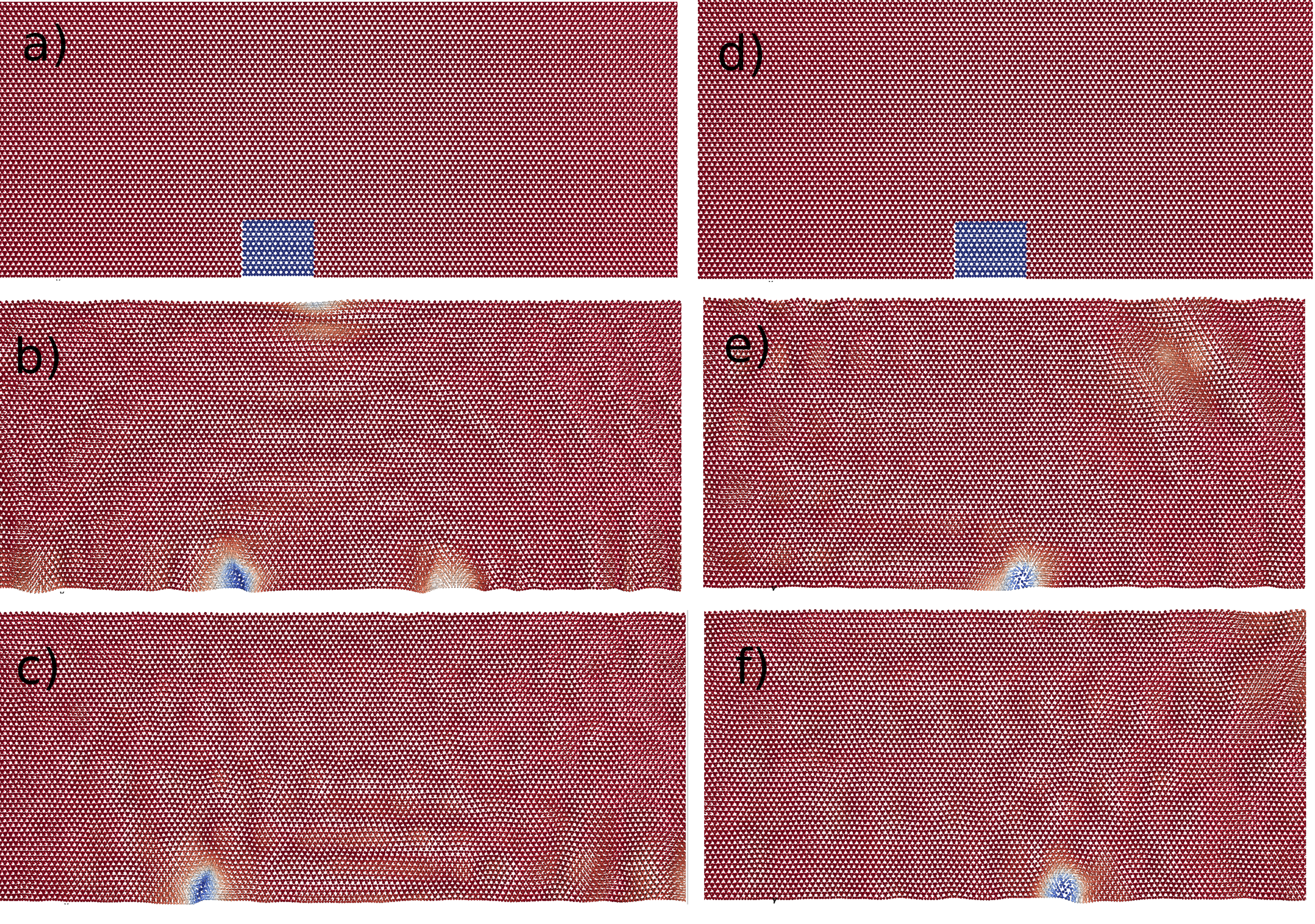}
 \caption{Snapshots of the meron (a-c) and antimeron (d-f) just after the excitation by an external local field. The exchange parameters are $J=1$~mRy and $D=0.1$~mRy. 
 }
 \label{figamm}
\end{figure}

Despite this, the values provided by Eq.~(\ref{skyrmion_number}) are still useful at the atomistic level and it is possible to extract the basic properties of the topological excitations. For example, the chirality of merons and antimerons can be determined by the sign of $\mathfrak{R}$. As shown in Fig.~\ref{fig6}(d), the skyrmion number is almost constant after approximately 5-10 ps of the meron/antimeron creation and remains without significant decay for at least 70 ps even though the DM parameter is reasonable small. This particularity makes a large family of magnetic materials with kagome crystal structure, to be candidates for a majority gate implementation.  Notice also that both merons and antimerons have almost the same absolute value of the skyrmion number ($\lvert\mathfrak{R}\rvert\approx 0.4$) in the regime where $\mathfrak{R}$ is almost constant. This is a property of topological excitations and thus, the calculated skyrmion number can be used to properly characterize merons and antimerons. In the proof of concept outlined in this article, when a meron and antimeron collide, they will annihilate leaving behind no topological structure and a vanishing skyrmion number. This property and the fact that these two structures move along different directions are the two principal properties needed for the formation of the majority gate. In Fig.~\ref{figamm}, snapshots just after the creation of the meron and antimeron are shown indicating the that each of them moves along different directions. (See also Supplementary Video 1-2).  


\begin{figure}[t]
 \centering
 \includegraphics[width=0.48\textwidth]{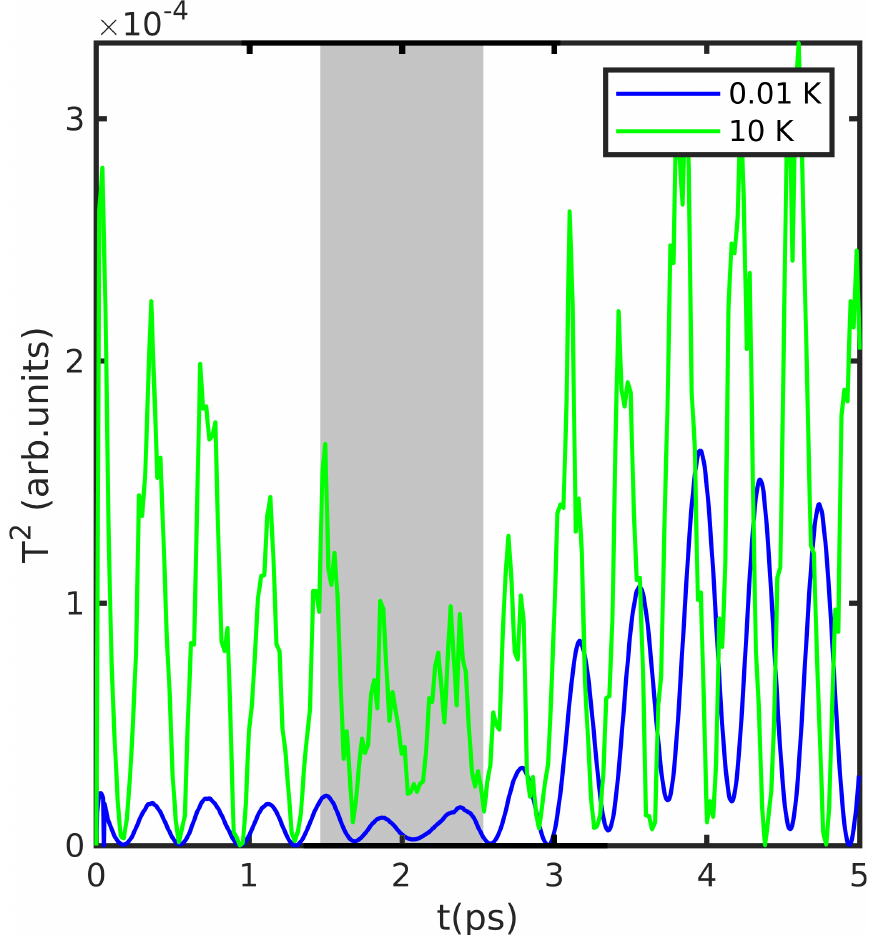}
 \caption{Torque squared as a function of the time for two different temperatures, i. e.,  $T=0.01$~K (blue) and $T=$10~K (green). Here $J=1$~mRy, $D=0.1$~mRy, $\alpha=0.01$, and external field equal to $10^4$~T.
}
 \label{fig8}
\end{figure}

A second important aspect to take into consideration for the efficiency of the majority gate device proposed here, is the stability under thermal fluctuations. For this reason, we also carried out simulations for the same system setup as described above, although with an increase of the temperature of the system. The thermal fluctuations are included via a stochastic field as it is described by Eq.~(\ref{stochastic}). Figure~\ref{fig8} shows the square of the torque given by Eq.~(\ref{torque1}), $T^2$, with respect to the excitation time (as in Fig.~\ref{fig3}) but for a temperature of $T=10$~K. For comparison, the same curve for $T=1$~mK is shown also in blue color. Due to the presence of thermal fluctuations, the $T^2$ signal is on average higher in intensity than the low temperature signal (cf. blue curve in Fig.~\ref{fig8}). The spiky  shape of the green curve around the peaks is evident and it is a consequence of the noise introduced by the thermal fluctuations. It is noticeable that the grey colored region, which is the critical region where meron and antimeron can be created, is almost identical with the one at $T=1$~mK. Even though the size of the grey region is almost the same as the one for $T=1$~mK, however the time interval needed for the application of the localized magnetic field in the rectangular region is slightly different. More specifically, for the  case with $T=10$~K, we have found a difference of 0.2 ps for the time needed in order to create a meron or an antimeron. Thus, in the temperature interval from 1~mK up to 10~K, the creation of the merons is almost unaffected. The latter holds for the smallest DM strength ($0.1$~mRy) tested in our simulations. For higher DM strength values ($D=0.25$~mRy), this threshold is increased up to 20~K. It is worthwhile to comment here that the aforementioned threshold-temperature is defined as the one where the $T^2$ curve follows the same trend as the curve for the case with very low temperature (~1 mK). It is key to determine the threshold-temperature since it provides a range in temperatures where the majority gate device outlined here is expected to work in a reliable way. Moreover, for the current values of $J$ and $D$ chosen in this work, the merons and antimerons can still be created up to 30K but the magnetic texture of the topological particles is substantially distorted and the excitation time is completely unpredictable, since the corresponding $T^2$ curve does not follow the curve at low temperatures. Note here that for higher values of the exchange parameters, so that the Curie Temperature of the material is much above room temperature, it is possible to stabilize merons and antimerons at 300 K~\cite{pereiro}.

\begin{figure}[t]
\includegraphics[width=0.48\textwidth]{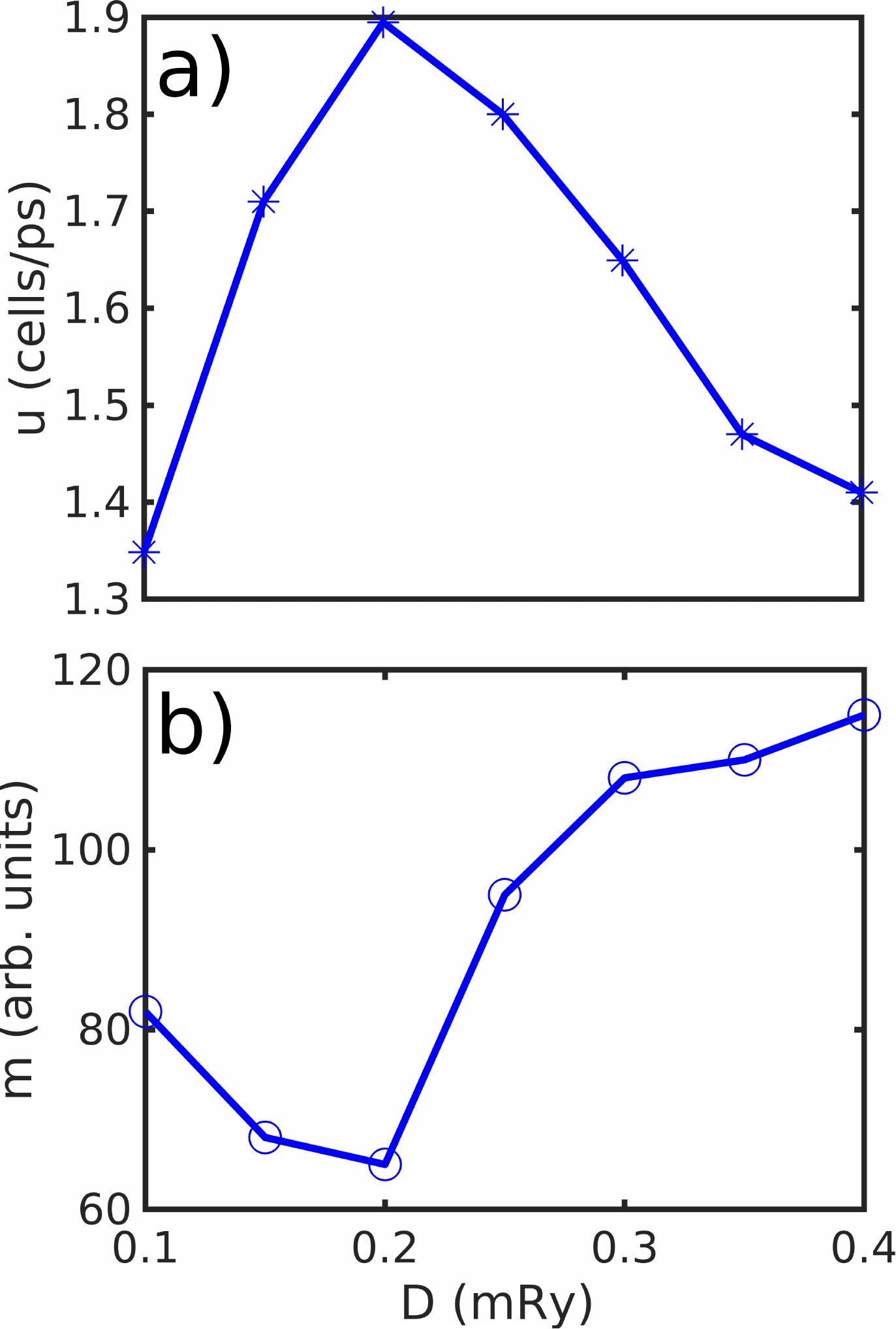}
\caption{Velocity (a) and size (b) of a meron with respect  to the strength D of the Dzyaloshinskii–Moriya interaction. The damping parameter was set to $\alpha=0.005$ while the external magnetic field used in the simulation is equal to $10^4$~T.}
\label{fig7}
\end{figure}

Another important property for the efficiency of a logic gate is the velocity of the carriers, as this aspect will rule the speed of the magnetic device itself. Figure~\ref{fig7}(a) shows the velocity and Fig.~\ref{fig7}(b) the mass (size) of the meron with respect to the strength of DM interaction. We define the size as the number of atomic spins for which the z component obeys the relation $|s_z/s|<0.5$, where $s$ and $s_z$ are the magnitude of the atomic magnetic moment and its z-component, respectively. It has been reported in Ref.~\cite{velocity} that the velocity of skyrmions is proportional to the DM strength but also to the size of the skyrmion. Our results do not fully comply with the trend found in Ref.~\cite{velocity}. Thus, in our findings, neither the velocity nor the size of the merons obey a monotonic behaviour with respect to the strength of the DM interaction. In Fig.~\ref{fig7} a), it is observed a change in the monotonic trend of the velocity in the neighborhood of $D=0.2$~mRy, so that the velocity increases up to a maximum value of 1.9 cells/ps and then it decreases down to roughly 1.4 cells/ps at $D=0.4$~mRy, whereas the mass follows the opposite trend. The reason for the discrepancy with the trend already proposed in Ref.~\cite{velocity} is most likely due to that the 
work in Ref.~\cite{velocity} was analyzing micromagnetic skyrmions rather than atomistic merons or antimerons, as studied here. Furthermore, they looked at current induced motion while we have another driving force.  Also, the uniqueness of the edge modes of the kagome lattice\cite{pereiro} provide couplings between skyrmion speed, mass and materials parameters that do not follow general trends.


\section{DEVICE FUNCTIONALITY}
\label{device}
\begin{figure}[h]
\centering
\includegraphics[width=0.48 \textwidth]{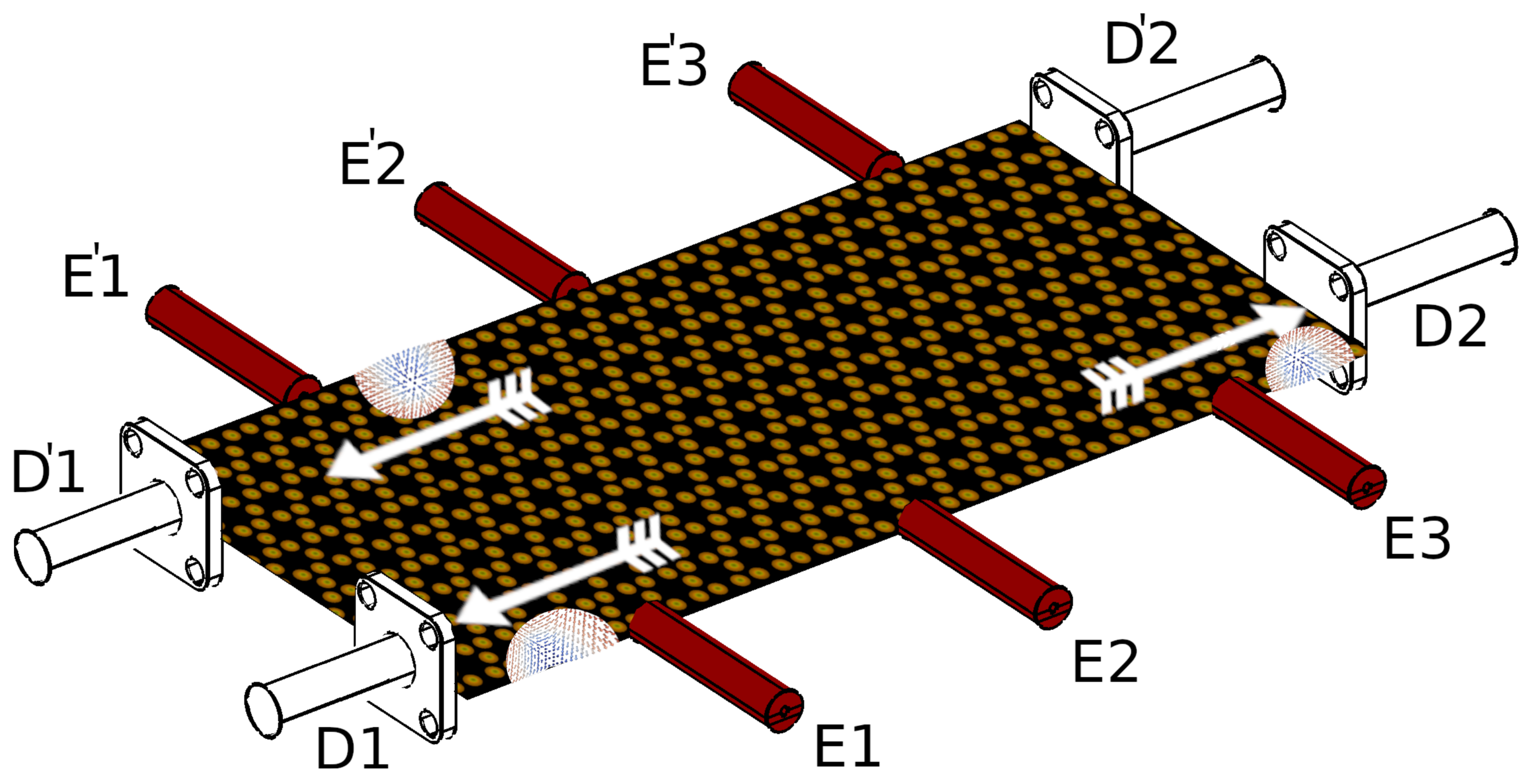}
\caption{Graphical diagram of the meron/antimeron gate. The black stripe represents the kagome stripe with atoms in yellow color. Red cylinders refer to the excitation regions, whereas the white ones represent the detectors. In this figure, a schematic of two gates denoted with and without primes on the same stripe are discussed in the text. In the example, a meron generated in E3 moves towards D2 and an antimeron generated in E1 moves clock-wise towards D1 while only a meron generated in E$^\prime$2 travels towards D$^\prime$1. The white arrows indicate the propagation direction of the pseudo-particles, i.e., clock-wise for antimerons and anticlock-wise for merons. }
\label{fig9}
\end{figure}

As already mentioned above, merons and antimerons follow a different direction when they travel along the edge of the kagome lattice. In the previous section, we described a way to control the creation of merons and antimerons by applying a local torque in a rectangular or circular tiny region. In the current section, we analyze a prototype of a magnetic majority gate based on merons and antimerons in a magnetic material with atomic planes in which the atoms arrange so as to form a kagome lattice. Figure~\ref{fig9} shows a schematic picture of the suggested device. 

In this graphical diagram, the red cylinders represent the excitation channels, through them a local torque is applied in order to  create the merons/antimerons, whereas the colorless ones represent the detectors (that rely on magneto resistance effects). The black stripe with yellow circles represents a stripe of the kagome lattice. The kagome stripe shown in Fig.~\ref{fig9} allows, by topological nature of the edge states in lattice, to build up two majority gates, each of them consisting from 3 input branches (initially we focus on E1-E3 
) and two detection branches (we focus first on D1-D2 
). It is worthwhile to mention here that we have added two detection points (D1 and D2) rather than just one, as it is the case for the canonical majority gate. 
Let us take an example were all excitations produce antimerons, at E1, E2 and E3. Then, all of them will move to the left, with the consequence that a signal will be measured only at D1. On the other hand, if all excitations produce merons, at E1, E2 and E3, they will move to the right and only the right branch (D2) will now detect a signal. Consequently, the presence of both detectors is mandatory in order to detect a majority signal. The two operations described above are two of the basic operations for the gate functionality. 
 
Next, we consider E1 as the branch that generates an antimeron while in E2 and E3 produce merons. Then, the antimeron will move to the left and the other two merons to right, thus detector D1 will measure an antimeron and D2 two merons.
For the sake of clarity, we represent an antimeron as bit 1 and a meron as bit 0. Supplementary videos 1 and 2 show the creation and motion of three antimerons and three merons respectively. Supplementary video 3 show the creation of two merons, one antimeron and their collision. 
In Table~\ref{table1}, we list all possible outcomes for the eight relevant input configurations, establishing a logical table. If the difference between D1 and D2 is positive, the majority input was an antimeron (1), which is listed under the column labelled 'majority signal'. If the difference between D1 and D2 is negative, the majority input was a meron (0). We reach similar conclusions as obtained in Ref.~\cite{1D_gate} that all situations in Table~\ref{table1} where E1 has entry 1 act as an OR gate for the signals E2 and E3, while all situations where E1 has entry 0 act as an AND gate for signals E2 and E3. This forms the basic functionality of the here suggested device and the temperature stability for the whole device operation is the same as for the creation/lifetime analysis shown above see (Supplementary video 4).

\begin{table}[t]
\centering
\caption{Logical table of the three-input non-canonical majority gate. For sake of clarity, we represent with 1 antimerons and with 0 merons. The output signals of detector D1 and D2 a listed for each case, and the difference in signals if the majority input was a meron (D1 - D2 $<$ 0) or antimeron (D1 - D2 $>$ 0).
}
\begin{tabularx}{\columnwidth}{ccccccc}
\hline
\hline\\[-0.30cm]
 In E1 & In E2 & In E3 & Out D1 & Out D2 & D1-D2 & majority signal\\
\hline
\\[-0.20cm]
 1 & 1 & 1 & 3 & 0 & + & 1\\
 1 & 1 & 0  & 2 & 1 & + & 1\\
 1 & 0 & 1 & 1 & 0 & + & 1\\
 1 & 0 & 0 & 1 & 2 & - & 0\\
\hline
 0 & 1 & 1 & 1 & 0 & + & 1\\
 0 & 0 & 1 & 0 & 1 & - & 0\\
 0 & 0 & 0 & 0 & 3 & - & 0\\
 0 & 1 & 0 & 0 & 1 & - & 0\\
\hline
\hline
\end{tabularx}
\label{table1}
\end{table}

In Fig.~\ref{fig9}, based on the concept developed above, two majority gates are presented on a single ferromagnetic kagome layer (inputs E1 - E3 for the first gate, and inputs E$^\prime$1 - E$^\prime$3 for the second gate). On the principle described in this section, only one single edge with no external guidance during operation is needed to perform a majority gate logical operation. Having this in mind, it becomes extremely tempting and promising to use the whole monolayer. As mentioned above, the vast majority of simulations were carried in a kagome monolayer with a length $l_y$ of 40 unit cells along the y direction. For this length scale, i.e., $l_y\approx40$~\AA~(assuming 1~\AA~unit cell dimension), the gate can have the same functionality along both edges without the one interfering with the other (See supplementary video 5). In consequence, one single kagome monolayer can be used to construct more than one majority gate. The concept of multiple gates in one atomic layer can be easily expanded, if the size of the monolayer increases to bypass the barrier of the number of cells required to avoid interference or coupling effects between the topological excitations belonging to different gates.


\section{DISCUSSION AND CONCLUSIONS}
\label{conclusions}
In this work, we have presented a basic idea on how to construct a magnetic majority gate on a single material. The kagome ferromagnet seems to be a promising solution due to the existance of edge modes. The latter allows for the existance of meron and antimeron solutions allowing for the direct assignment of bits in different topologies surviving in the same material at the same time.

Regarding real materials where the proof of concept proposed here can be realized in practice, several proposals exists like namely vanadate pyrochlores with generic formula A$_2$V$_2$O$_7$ with A=Lu, Yb, Tm, Y, SrCr$_8$Ga$_4$O$_{19}$, Ba$_2$Sn$_2$Ga$_3$ZnCr$_7$O$_22$ and magnetic jarosites like KM$_3$(OH)$_6$(SO$_4$)$_2$ with M=V, Cr, or Fe. Also Fe$_3$Sn$_2$ might be a suitabale solution as in this material, Fe ions form bilayers of offset kagome networks, while the Sn ions occupy two distinct crystallographic sites and importantly this material has a Curie temperature close to 640~K making it a really strong candidate for even room temperature operations. In fact, in Ref.~\cite{hou} it has been experimentally observed magnetic skyrmionic bubbles in Fe$_3$Sn$_2$ at room temperature as it was anticipated and predicted in Ref.~\cite{pereiro}. Another promising  material is CrI$_3$ which is a honeycomb ferromagnet. Magnetism in CrI$_3$ is assigned to  Cr$^3+$ anions surrounded by I$_6$ octahedra, forming a 2D honeycomb network. Ferromagnetic ordering appears with Cr$^3+$ spins oriented along the c axis providing a conditions for DM interactions to open spin wave gaps~\cite{Chrome1,pereiro}. For example, the experimental magnon spectra measured in Ref.~\cite{Chrome2} exhibit remarkably large gaps at the Dirac points indicating the presence of DM interactions. As the gap is induced from DM, CrI$_3$ should have topological edge states and, in consequence, it would be suitable for spintronic applications. 

In this work, the external torque is created by an external local magnetic field. The latter is not the most efficient approach as the local field to break magnetic order is quite large. Notice that a huge local external field is clearly physical as it needs to overcome the exchange energy to reverse a small region of the material and thus, leading to high power consumption for the field generation. Probably an application of laser pulse to locally heat up the stripe will lead to more efficient approach.
Notably though with proper selection of the field application time, we can control the creation of the specific topologies, leading to a robust  operation of the gate. 

By taking advantage of the edge modes present in the kagome magnet, the created merons and antimerons do not need any external guidance of force, as they move by the aid of the topologican nature of the magnon edge modes\cite{pereiro}. A really versatile outcome of this approach is that the operation of the gate needs only one edge of the kagome 2D stripe. As a result, it is possible to build multiple gates even in one single monolayer by making use of the different edges of the monolayer. In terms of the material, the only constrain is the size of the monolayer which has to be chosen with dimensions as large as there is no edge-to-edge interaction. The 40 cells length chosen here along the y-direction, it is a safe criterion for that threshold. The second constrain is related to the detection system. 

According to our approach, the gate does not operate as a conventional majority, but in contrast, the detection system consists always by two initial detectors which are able to identify merons or antimerons. Thus, in a multiple gate scenario, a large number of detectors need to be used. In the scenario of the minimum number of detectors, i.e. 4 detectors (cf.~\ref{fig9}) the gates should operate in sequential form under different time windows in order for the detectors to receive signals from a specific gate.

The detection system is a critical part for efficient and practical spintronic applications.   
Recent works show that magnetic tunnel junction, which can be constructed by lithography, can be used to identify single skyrmions~\cite{detect1,detect2}. Also in Ref.~\cite{detect3}, the authors propose a clearly electrical detection of the position  of skyrmions in conducting systems based on the Hall resistance. For the three-input majority device advanced in this article, the setup for the detection system might be even simpler than in previous published prototypes \cite{detect3}. As already commented, merons and antimerons leave only on the edge of stripe and are constrained to move along specific directions,i.e., the merons always move counterclockwise while antimerons travel clockwise. Thus, detectors D1 and D2 always measure antimerons and merons, respectively. In such a scenario, the detectors only need to identify the particle passing by near them without caring about the atomic detail of the magnetic texture. Consequently, a magnetoresistive sensor  might be enough to ensure a proper detection of the topological excitations.

\begin{center}
{\bf ACKNOWLEDGMENTS}
\end{center}

The authors acknowledge financial support from Knut and Alice Wallenberg Foundation through Grant No. 2018.0060. A.D. acknowledges financial support from the Swedish Research Council (VR) through Grants No. 2015-04608, No. 2016-05980, and No. 2019-05304. O.E. also acknowledges support from eSSENCE, SNIC, the Swedish Research Council (VR) and the ERC (synergy grant). D.T. acknowledges support from the Swedish Research Council (VR) through Grant No. 2019-03666. E.S. acknowledges financial support from the Swedish Research Council (VR) through Grant No. 2017-03832.
Some of the computations were performed on resources provided by the Swedish National Infrastructure for Computing (SNIC) at the National Supercomputer Center (NSC), Linköping University, the PDC Centre for High Performance Computing (PDC-HPC), KTH, and the High Performance Computing Center North (HPC2N), Umeå University.

\bibliographystyle{apsrev4-2}

\begin{thebibliography}{49}
\bibitem{majority} In a 3-input majority gate, the minority of the inputs is only one single excitation.
\end{thebibliography}

\end{document}